\documentstyle[amssymb,aps,prb,twocolumn]{revtex}
\begin{document}
\title{Kramers equation for a charged Brownian particle: The exact solution }
\author{Tania P. Sim\~{o}es$^{a)}$ and Roberto E. Lagos$^{b),*}$}
\address{$^{a)}$ Instituto de F\'{\i }sica `{\it Gleb Wataghin'} Universidade
Estadual de Campinas \\
(UNICAMP) CP. $6165$, $13083-970$ Campinas, SP Brazil.\\
$^{b)}$ Departamento de F\'{\i }sica, IGCE Universidade Estadual Paulista\\
(UNESP) CP. $178$, $13500$-$970$ Rio Claro, SP, Brazil.\\
$^{*}$ Virtual Research Center -CC@Complex -Science and Computing\\
for Complexity (UNESP SP\ Brazil)}
\maketitle

\begin{abstract}
We report the exact fundamental solution for Kramers equation associated to
a brownian gas of charged particles, under the influence of homogeneous
(spatially uniform) otherwise arbitrary, external mechanical, electrical and
magnetic fields. Some applications are presented, namely the
hydrothermodynamical picture for Brownian motion in the long time regime.
\end{abstract}

\pacs{05.20.Dd, 05.40.Jc,51.10+y.}

%\date{\today}

\narrowtext

\vspace{0.5cm}

\section{Introduction}

In Chandrasekhar's 1943 celebrated paper\cite{chandra}, Kramers equation 
\cite{kram} was solved for the free Brownian particle and some general lines
were drawn towards solving this problem in a field of force. Only recently
some progress was reported considering Kramers equation for a charged
Brownian particle in a field of force: Czopnik \& Garbaczewsky (CG) \cite
{garba} solved Kramers planar equation in a magnetic field, essentially
transforming the magnetic field contribution into a tensorial Stokes-like
dissipative force. Later, Ferrari \cite{ferrari} via transformed phase space
variables, mapped Kramers equation for a charged Brownian particle in an
electric field into the free Brownian particle case. By combining both CG's
`rotated' Stokes force and Ferrari's gauge, in\ section II we report the
exact fundamental solution of Kramers equation for a charged Brownian
particle in an uniform, otherwise arbitrary field of forces. In section III
we present some applications, concerning the hydrothermodynamical picture of
Brownian motion, the validity of the local equilibrium approximation and the
`linear' regime (see for example \cite{jou1,rubi}). Comparison is made with
some results obtained via a perturbative recursive scheme \cite{bcl1,bcl2}.
Finally in section IV we present some concluding remarks and outline some
work in progress.

\section{Fundamental solution}

We study a Brownian gas composed of charged particles (mass $m$, charge $e$)
under the influence of external fields: mechanic (${\bf mec}$), electric ($%
{\bf E}$) and magnetic (${\bf B}$) fields, uniform in space and in general
time dependent. Our starting point is {\bf Kramers} equation for the density
probability distribution $P({\bf x},{\bf v,}t)$ in phase space (position $%
{\bf r}$, velocity ${\bf v}$) at time $t$, in contact with a reservoir at
temperature $T_{R}$ and under the force fields

\begin{equation}
{\bf F(v,}t{\bf )=F}_{mec}+e{\bf E}+\frac{e}{c}{\bf v\times B}  \label{force}
\end{equation}

\noindent The associated Kramers equation \cite{kram} reads

\begin{equation}
\frac{\partial P}{\partial t}+{\bf v}\frac{\partial P}{\partial {\bf x}}+%
\frac{{\bf F}}{m}\frac{\partial P}{\partial {\bf v}}=-\frac{\partial }{%
\partial {\bf v}}\frac{{\bf F}_{d}}{m}P+\lambda \frac{T_{R}}{m}\frac{%
\partial ^{2}P}{\partial {\bf v}^{2}}  \label{original}
\end{equation}

\noindent with Boltzmann\'{}s constant $k_{B}\equiv 1,$ $\lambda =\tau ^{-1}$
is the friction coefficient (inverse of the collision time) and ${\bf F}%
_{d}=-\lambda m{\bf v}$ is the dissipative Stokes-like force. As in previous
work \cite{bcl1,bcl2}, where a perturbative recursive scheme was presented
for a more general case, we define $v_{T}$, a thermal velocity given by $%
mv_{T}^{2}=T_{R}$ and dimensionless variables by scaling space, velocity and
time, respectively with $l=\tau v_{T},$ $v_{T}$ and $\tau $. Also we define
the conservative acceleration ${\bf a}$ (and the associated potential $\phi $%
) and the cyclotronic frequency vector ${\bf \Omega }$ respectively as
(hereafter all quantities are dimensionless unless stated otherwise)

\begin{eqnarray}
{\bf a} &=&\frac{\tau }{mv_{T}}\left( {\bf F}_{mec}+e{\bf E}\right) =-\frac{%
\partial \phi }{\partial {\bf x}}  \label{acel} \\
{\bf \Omega } &=&\frac{e\tau }{mc}{\bf B=}\omega \stackrel{\wedge }{\bf %
\omega }
\end{eqnarray}

\noindent Notice that in terms of the usual dimensional cyclotronic
frequency \cite{bcl2}, we have $\omega =\omega _{c}\tau $. Concerning
notation, we chose the very convenient bra-ket convention. Denote any vector 
${\bf V}$ by ${\bf V=}V_{x}\left| x\right\rangle +V_{y}\left| y\right\rangle
+V_{z}\left| z\right\rangle $ and its adjoint by ${\bf V}^{\dagger
}=V_{x}\left\langle x\right| +V_{y}\left\langle y\right| +V_{z}\left\langle
z\right| $ (all quantities are real). We also define some useful dyadics,
namely: ${\bf e}_{1}=\left| z\right\rangle \left\langle z\right| $, ${\bf e}%
_{2}=\left| x\right\rangle \left\langle x\right| +\left| y\right\rangle
\left\langle y\right| $, ${\bf e}_{3}=\left| x\right\rangle \left\langle
y\right| -\left| y\right\rangle \left\langle x\right| $ and the unit dyadic $%
{\bf e=e}_{1}+{\bf e}_{2}$. Furthermore, we define the z-axis as the
magnetic field direction ($\stackrel{\wedge }{\bf \omega }=\stackrel{\wedge 
}{z}$), and Kramers equation (\ref{original}) is cast in a compact form as

\begin{equation}
\frac{\partial P}{\partial t}+{\bf v}\frac{\partial P}{\partial {\bf x}}+%
{\bf a}\frac{\partial P}{\partial {\bf v}}=\frac{\partial }{\partial {\bf v}}%
{\bf \Lambda v}P+\frac{\partial ^{2}P}{\partial {\bf v}^{2}}  \label{kramers}
\end{equation}

\noindent where the magnetic contribution is included as a tensorial
Stokes-like dissipative term (see \cite{garba,bcl2})

\[
{\bf \Lambda v=v+\Omega \times v=}({\bf e}-\omega {\bf e}_{3}){\bf v} 
\]

\noindent We also define

\[
{\bf M=\Lambda }^{-1}{\bf =e}_{1}+\alpha \left( {\bf e}_{2}+\omega {\bf e}%
_{3}\right) 
\]

\noindent with $\alpha ^{-1}=1+\omega ^{2}$. In dimensional form this
tensorial collision time constant has the familiar form \cite{bcl2}

\begin{equation}
{\bf M=}\frac{\tau }{1+(\omega _{c}\tau )^{2}}\left( 
\begin{array}{ccc}
1 & \omega _{c}\tau & 0 \\ 
-\omega _{c}\tau & 1 & 0 \\ 
0 & 0 & 1+(\omega _{c}\tau )^{2}
\end{array}
\right)  \label{expansion}
\end{equation}

Since the planar ($x,y$) dynamics is decoupled from the $z$ axis dynamics,
the case ${\bf a\equiv 0}$ can be trivially solved generalizing the planar
results of (CG) \cite{garba}. Then, transforming to new variables ${\bf R}$
and ${\bf V}$, via Ferrari's gauge \cite{ferrari} we map our problem to the
free Brownian particle solved in 1943 by Chandrasekhar \cite{chandra}. Here
we particularize to time independent external fields case, yielding
Ferrari's transformed variables \cite{ferrari}

\begin{eqnarray*}
{\bf R} &=&{\bf x-\delta x(a,M,}t) \\
{\bf V} &=&{\bf v-\delta v(a,M,}t)
\end{eqnarray*}

\noindent where

\begin{eqnarray*}
{\bf \delta v} &=&{\bf M(1-\Theta )a+\Theta v}_{0} \\
{\bf \delta x} &=&{\bf Ma}t-{\bf M}^{2}\left( 1-{\bf \Theta }\right) {\bf a}+%
{\bf M}\left( 1-{\bf \Theta }\right) {\bf v}_{0}+{\bf x}_{0}
\end{eqnarray*}

\[
{\bf \Theta }=\exp (-t){\bf (e}_{1}+{\bf e}_{2}\cos \omega t+{\bf e}_{3}\sin
\omega t{\bf )} 
\]

\noindent and with Kramers equation (\ref{kramers}) mapped to the trivially
solved \cite{garba,ferrari,chandra} equation

\[
\frac{\partial P}{\partial t}+{\bf V}\frac{\partial P}{\partial {\bf R}}=%
\frac{\partial }{\partial {\bf V}}{\bf \Lambda V}P+\frac{\partial ^{2}P}{%
\partial {\bf V}^{2}} 
\]

\noindent The fundamental solution $G({\bf x,t,}t|{\bf x}_{0},{\bf v}_{0}),$
namely with free boundary conditions and the point-like initial condition

\[
G({\bf x},{\bf v,}t=0|{\bf x}_{0},{\bf v}_{0})=\delta ({\bf x-x}_{0})\delta (%
{\bf v-v}_{0}) 
\]

\noindent is given by

\begin{equation}
G({\bf x,t,}t|{\bf x}_{0},{\bf v}_{0})=\left( \frac{1}{2\pi }\right) ^{3}%
\frac{1}{\Delta \sqrt{\Delta ^{*}}}\exp -\frac{1}{2}{\bf \Phi }
\label{green}
\end{equation}

\noindent where

\[
{\bf \Phi }={\bf \Phi }_{d}-{\bf \Phi }_{s}-{\bf \Phi }_{a} 
\]

\noindent with

\begin{eqnarray*}
{\bf \Phi }_{d} &=&{\bf V}^{\dagger }{\bf A}_{v}{\bf V}+{\bf R}^{\dagger }%
{\bf A}_{r}{\bf R} \\
{\bf \Phi }_{sn} &=&{\bf V}^{\dagger }{\bf A}_{m}{\bf R+R}^{\dagger }{\bf A}%
_{m}{\bf V} \\
{\bf \Phi }_{a} &=&2{\bf Q}^{\dagger }\left( {\bf R\times V}\right)
\end{eqnarray*}

\noindent or in a compact form

\[
{\bf \Phi }=\left( 
\begin{array}{cc}
{\bf V}^{\dagger } & {\bf R}^{\dagger }
\end{array}
\right) {\bf A}\left( 
\begin{array}{c}
{\bf V} \\ 
{\bf R}
\end{array}
\right) 
\]

\noindent where

\[
{\bf A}=\left( 
\begin{array}{cc}
{\bf A}_{v} & {\bf -A}_{m}+\left| {\bf Q}\right| {\bf e}_{3} \\ 
-{\bf A}_{m}-\left| {\bf Q}\right| {\bf e}_{3} & {\bf A}_{r}
\end{array}
\right) 
\]

\noindent with

\begin{equation}
{\bf A}_{\alpha }=\frac{a_{\alpha }}{\Delta }{\bf e}_{2}+\frac{a_{\alpha
}^{*}}{\Delta ^{*}}{\bf e}_{1}  \label{coeff}
\end{equation}

\begin{eqnarray*}
{\bf Q} &=&\frac{k}{\Delta }\stackrel{\wedge }{\bf \omega } \\
\hspace{1cm}\Delta &=&a_{v}a_{r}-a_{m}^{2}-k^{2}
\end{eqnarray*}

\noindent and where

\[
a_{r}=1-b_{e}^{2} 
\]

\[
a_{m}=\alpha \left( 1-2b_{e}b_{c}+b_{e}^{2}\right) 
\]

\[
k=\alpha \left( 2b_{e}b_{s}-\omega a_{r}\right) 
\]

\[
a_{v}=\alpha \left( a_{r}+2t-4\alpha \left( 1+b_{e}\left( \omega
b_{s}-b_{c}\right) \right) \right) 
\]

\[
b_{e}=\exp (-t),\hspace{0.5cm}b_{c}=\cos \omega t,\hspace{0.5cm}b_{s}=\sin
\omega t 
\]

\noindent The superscript $*$ in equations (\ref{green}, \ref{coeff})
denotes the corresponding quantity evaluated at null magnetic field $(\omega
\equiv 0).$

The general solution satisfying the initial condition $P({\bf x},{\bf v,}%
t=0)=P_{0}({\bf x,v)}$ is given by

\begin{equation}
P{\bf (x,v,}t){\bf =}\int d{\bf x}_{0}d{\bf v}_{0}G({\bf x,v,}t|{\bf x}_{0},%
{\bf v}_{0})P_{0}({\bf x}_{0}{\bf ,v}_{0}{\bf )}  \label{general}
\end{equation}

\noindent We normalize the probability $P$ to $N$, the total number of
particles in the gas and define the particle density

\[
n({\bf x,}t)=\int d{\bf v}P{\bf (x,v,}t{\bf )} 
\]

\noindent requiring

\[
N=\int d{\bf x}d{\bf v}P_{0}({\bf x,v)}=\int d{\bf x}d{\bf v}P({\bf x,v},t%
{\bf )} 
\]

\noindent A mean density $n_{0}$ is defined in the thermodynamic limit as $%
N=n_{0}V$ where the volume is defined by $V=\int d{\bf x}$. Thermal
equilibrium conditions (TEC) are reached in the asymptotic regime $%
t\rightarrow \infty $ under null external fields, where we retrieve the
Maxwellian distribution

\[
P_{TEC}{\bf (v}){\bf =}\left( 2\pi \right) ^{-3/2}n_{0}\exp -\frac{1}{2}{\bf %
v}^{2} 
\]

\section{Hydrothermodynamics of Brownian motion}

We start by defining some quantities of interest \cite{bcl1,bcl2}: particle
flow density ${\bf J}$ , the associated hydrodynamic velocity ${\bf u}$,
kinetic energy density $\varepsilon $, local gas temperature $\theta $ (in $%
T_{R}$ units), local pressure $p$, entropy density $s$, Gibbs energy density 
$g$ and the total chemical potential $\mu $, respectively defined by

\begin{equation}
{\bf J}({\bf x,}t)=\int d{\bf vv}P({\bf x},{\bf v,}t)=n({\bf x,}t){\bf u}(%
{\bf x,}t)  \label{flow1}
\end{equation}

\[
\varepsilon ({\bf x,}t)=\frac{3}{2}n({\bf x,}t)\theta ({\bf x,}t)=\frac{1}{2}%
\int d{\bf vv}^{2}P({\bf x},{\bf v,}t) 
\]

\[
p({\bf x,}t)=n({\bf x,}t)\theta ({\bf x,}t) 
\]

\[
s({\bf x,}t)=-\int d{\bf v}P({\bf x,v},t)\ln \varkappa P({\bf x,v},t) 
\]

\[
g({\bf x},t)=\varepsilon ({\bf x},t)-\theta ({\bf x},t)s({\bf x},t)+p({\bf x}%
,t) 
\]

\[
\mu ({\bf x},t)=\frac{g({\bf x},t)}{n({\bf x},t)}+\phi 
\]

\noindent The additive constant

\[
\ln \varkappa =-1+3\ln \frac{h}{\tau T_{R}} 
\]

\noindent is chosen such that under TEC\ we retrieve the usual
thermodynamical entropy density \cite{bcl1,bcl2}. This equilibrium entropy
and the associated chemical potential are respectively given by (in
dimensional form)

\[
s_{eq}(n_{0},T_{R})=n_{0}\left( \frac{5}{2}+\ln \frac{n_{Q}\left(
T_{R}\right) }{n_{0}}\right) 
\]

\[
\mu _{eq}\left( n_{0},T_{R}\right) =-T_{R}\ln \frac{n_{Q}\left( T_{R}\right) 
}{n_{0}} 
\]

\noindent with

\[
n_{Q}\left( T_{R}\right) =\left( \frac{2\pi mT_{R}}{h^{2}}\right) ^{\frac{3}{%
2}} 
\]

\noindent Furthermore we define an entropy flux density as

\[
{\bf J}_{s}=-\int d{\bf vv}P({\bf x,v},t)\ln \varkappa P({\bf x,v},t) 
\]

Balance equations are \ computed as in \cite{bcl1,bcl2}, yielding the
continuity (Smoluchowsky) and the entropy balance equations, given
respectively by

\[
\frac{\partial n}{\partial t}+{\bf \nabla J=}0 
\]

\noindent and

\[
\frac{\partial s({\bf x,}t)}{\partial t}+\frac{\partial {\bf J}_{s}({\bf x,}%
t)}{\partial {\bf x}}=\sigma ({\bf x,}t) 
\]

\noindent where the entropy production density $\sigma ({\bf x,}t)$ is given
by

\[
\sigma ({\bf x,}t)=\int d{\bf v}P({\bf x,v},t)\left( \frac{\partial \ln P(%
{\bf x,v},t)}{\partial {\bf v}}\right) ^{2}-3n({\bf x,}t) 
\]

In this brief report we compute the non equilibrium long time regime (LTR,
the limit $t\gg 1$ and with non zero external fields). In this limit we have:

\begin{eqnarray*}
\delta {\bf v} &=&{\bf -M}\frac{\partial \phi }{\partial {\bf x}}={\bf V}_{F}
\\
\delta {\bf x} &=&{\bf V}_{F}t+{\bf x}^{*}\hspace{0.5cm}{\bf x}^{*}={\bf x}%
_{0}-{\bf M(V}_{F}-{\bf v}_{0}{\bf )}
\end{eqnarray*}

\noindent Thus, in the LTR Ferrari's \cite{ferrari} transformation is simply
represented by the constant velocity shift ${\bf V}_{F}$ or equivalently,
given by the solution of the (dimensional) equation

\[
\frac{{\bf V}_{F}}{\tau }{\bf =}\frac{1}{m}\left( {\bf F}_{mec}+e{\bf E}+%
\frac{e}{c}{\bf V}_{F}{\bf \times B}\right) 
\]

We highlight some results in the LTR, corroborating some previously obtained
results, via a recursive method \cite{bcl1,bcl2}. We may cast

\[
P({\bf x},{\bf v},t)=n({\bf x},t)W({\bf x},{\bf v},t) 
\]

\noindent where to lowest order in $t^{-1}$ we have

\[
W({\bf x},{\bf v},t)=\left( \frac{1+2t}{4\pi t}\right) ^{\frac{3}{2}}\exp -%
\frac{(1+2t)}{4t}\left( {\bf V-V}_{D}\right) ^{2} 
\]

\[
n({\bf x},t)=\left( \frac{1}{4\pi t}\right) ^{\frac{3}{2}}\frac{1}{\alpha }%
\exp -\frac{1}{2}\Gamma ({\bf R}) 
\]

\noindent with

\[
\Gamma ({\bf R})=\frac{1}{2\alpha t}\left( {\bf R}^{2}+\alpha \left( {\bf %
\Omega R}\right) ^{2}\right) -{\bf V}_{D}^{2} 
\]

\[
{\bf V}_{D}=\frac{1}{2t}\left( {\bf R+R\times \Omega }\right) 
\]

Then it is also satisfied the relation 
\begin{equation}
{\bf J}({\bf x},t)=-{\bf M}\left( \frac{\partial n}{\partial {\bf x}}+n\frac{%
\partial \phi }{\partial {\bf x}}\right)  \label{flow}
\end{equation}

\noindent We define a magneto covariant derivative as in \cite{bcl2} for any
given function $f({\bf x},t)$ as (notice the potential $\phi $ is expressed
in $T_{R}$ units)

\begin{eqnarray*}
D_{x}f &=&{\bf M}\exp (-\phi )\frac{\partial }{\partial {\bf x}}f({\bf x,}%
t)\exp (\phi ) \\
&=&{\bf M}\left( \frac{\partial f({\bf x,}t)}{\partial {\bf x}}+f({\bf x,}t)%
\frac{\partial \phi }{\partial {\bf x}}\right)
\end{eqnarray*}

\noindent and from (\ref{flow1}) and (\ref{flow}) we retrieve Smoluchowsky
equation with a magnetic field \cite{bcl2}

\begin{equation}
\frac{\partial n}{\partial t}=\frac{\partial }{\partial {\bf x}}\left(
D_{x}n\right)  \label{marian}
\end{equation}

Also, the local temperature $T({\bf x},t)$, the entropy density and the
total chemical potential (in dimensional form) are given by the following
expressions

\begin{equation}
\frac{T({\bf x},t)}{T_{R}}=\theta ({\bf x},t)=1-\frac{1}{2t}+\frac{1}{3}{\bf %
u}^{2}({\bf x,}t)  \label{temp}
\end{equation}

\begin{equation}
s({\bf x},t)=s_{eq}\left( n({\bf x,}t),T({\bf x,}t)\right) +\delta s({\bf x,}%
t)  \label{entropy}
\end{equation}

\begin{equation}
\mu ({\bf x},t)=\mu _{eq}\left( n({\bf x,}t),T({\bf x,}t)\right) +\delta \mu
({\bf x},t)  \label{chemical}
\end{equation}

\noindent where the entropy and chemical potential shifts, are respectively
given by \noindent (to lowest order in $t^{-1}$ and ${\bf u}^{2}\sim {\bf M}%
^{2}$)

\[
\delta s({\bf x,}t)=-\frac{1}{2}n({\bf x,}t){\bf u}^{2}({\bf x,}t)\hspace{%
0.5cm} 
\]

\[
\delta \mu ({\bf x},t)=\frac{1}{2}T({\bf x,}t){\bf u}^{2}({\bf x,}t)+\phi (%
{\bf x,}t) 
\]

\noindent Furthermore, the hydrodynamic velocity ${\bf u}$ may be cast in
several equivalent forms, namely

\[
{\bf u}({\bf x,}t)=D_{x}\ln n={\bf M}\frac{\partial }{\partial {\bf x}}%
\left( \frac{\mu ({\bf x,}t)}{\theta ({\bf x,}t)}\right) ={\bf V}_{D}+{\bf V}%
_{F} 
\]

\noindent Lat us briefly comment on the last equation. The entropy density
(equation \ref{entropy}) with ${\bf u}({\bf x,}t)=D_{x}\ln n$ is formally
reminiscent of a Ginzburg-Landau expansion as noted in \cite{bcl2}. Also, as
stated by Landauer \cite{landa} the gradient of a bona fide nonequilbrium
chemical must be proportional to the particle flux ${\bf J}=n{\bf u}$.

It was found \cite{bcl2} that a suitable expansion parameter is the
collision time $\tau $ (or its tensorial partner ${\bf M}$). In the LTR and
confirming our conclusions in \cite{bcl2}, for the Brownian motion of a
charged particle, the local equilibrium hypothesis is satisfied {\em only to
first order} (linear) in $\tau ({\bf M})$ and where, among other things,
Onsager relations are satisfied. In this linear approximation we have $T(%
{\bf x},t)\approx T_{R},$ $\delta s({\bf x,}t)\approx 0,\delta \mu ({\bf x,}%
t)\approx \phi ({\bf x,}t).$ Furthermore up to second order in $\tau $

\[
{\bf J}({\bf x,}t)\approx \frac{1}{2}\frac{\partial \phi }{\partial {\bf x}}%
\hspace{0.5cm}\sigma ({\bf x,}t)\approx \frac{3}{2t}n({\bf x,}t) 
\]

\noindent yielding a nonequilibrium temperature independent of the magnetic
field, and a positive definite entropy production that vanishes as $t^{-1}.$

\section{Concluding remarks}

Here we have presented the fundamental exact solution for Kramers equation
in a field of uniform forces, hitherto unknown, and applied the results to
the long time regime. Work in progress seeks solutions for general initial
conditions and other than free boundary conditions (for example membranes 
\cite{membr}) and the associated hydrothermodynamical picture, to be
inserted into a more general framework\cite{jou1,rubi}. Also, we will
address the question of a nonuniform reservoir temperature $T_{R}({\bf x})$ 
\cite{bcl2,jou2} and the inclusion of chemical reactions\cite{bcl2,tania} to
our Brownian scheme.

\noindent {\em Acknowledgment}: This work was partially supported by CNPq
(Brasil), and submitted in partial fulfillment of the requirements for the
doctoral degree (TPS).

\end{document}